\documentclass[twocolumn]{aastex631}

\submitjournal{ApJL}

\begin{document}

\title{PHANGS-JWST First Results: Spurring on Star Formation:\\JWST Reveals Localised Star Formation in a Spiral Arm Spur of NGC~628}

\newcommand{\mpia}{Max-Planck-Institut f\"{u}r Astronomie, K\"{o}nigstuhl 17, D-69117, Heidelberg, Germany}
\newcommand{\oxford}{Sub-department of Astrophysics, Department of Physics, University of Oxford, Keble Road, Oxford OX1 3RH, UK}
\newcommand{\mcmaster}{Department of Physics and Astronomy, McMaster University, Hamilton, ON L8S 4M1, Canada}
\newcommand{\cita}{Canadian Institute for Theoretical Astrophysics (CITA), University of Toronto, 60 St George Street, Toronto, ON M5S 3H8, Canada}
\newcommand{\ghent}{Sterrenkundig Observatorium, Universiteit Gent, Krijgslaan 281 S9, B-9000 Gent, Belgium}
\newcommand{\oan}{Observatorio Astron\'{o}mico Nacional (IGN), C/Alfonso XII, 3, E-28014 Madrid, Spain}
\newcommand{\ari}{Astronomisches Rechen-Institut, Zentrum f\"{u}r Astronomie der Universit\"{a}t Heidelberg, M\"{o}nchhofstra\ss e 12-14, D-69120 Heidelberg, Germany}
\newcommand{\anu}{Research School of Astronomy and Astrophysics, Australian National University, Canberra, ACT 2611, Australia}   
\newcommand{\ita}{Universit\"{a}t Heidelberg, Zentrum f\"{u}r Astronomie, Institut f\"{u}r theoretische Astrophysik, Albert-Ueberle-Str. 2, 69120 Heidelberg, Germany}
\newcommand{\mpe}{Max-Planck-Institut f\"{u}r extraterrestrische Physik, Giessenbachstra{\ss}e 1, D-85748 Garching, Germany}
\newcommand{\cool}{Cosmic Origins Of Life (COOL) Research DAO, coolresearch.io}
\newcommand{\manch}{Jodrell Bank Centre for Astrophysics, Department of Physics and Astronomy, University of Manchester, Oxford Road, Manchester M13 9PL, UK}
\newcommand{\eso}{European Southern Observatory, Karl-Schwarzschild-Stra{\ss}e 2, 85748 Garching, Germany}
\newcommand{\cral}{Univ Lyon, Univ Lyon1, ENS de Lyon, CNRS, Centre de Recherche Astrophysique de Lyon UMR5574, F-69230 Saint-Genis-Laval France}
\newcommand{\cfa}{Center for Astrophysics, Harvard \& Smithsonian, 60 Garden St, Cambridge, MA, United States}
\newcommand{\CCAPP}{Center for Cosmology and Astroparticle Physics, 191 West Woodruff Avenue, Columbus, OH 43210, USA}
\newcommand{\OSU}{Department of Astronomy, The Ohio State University, 140 West 18th Avenue, Columbus, Ohio 43210, USA}
\newcommand{\iwr}{Universit\"{a}t Heidelberg, Interdisziplin\"{a}res Zentrum f\"{u}r Wissenschaftliches Rechnen, Im Neuenheimer Feld 205, D-69120 Heidelberg, Germany}
\newcommand{\UBonn}{Argelander-Institut f\"{u}r Astronomie, Universit\"{a}t Bonn, Auf dem H\"{u}gel 71, 53121, Bonn, Germany}
\newcommand{\stsci}{Space Telescope Science Institute, 3700 San Martin Drive, Baltimore, MD, USA}

\suppressAffiliations

\correspondingauthor{Thomas G. Williams}
\email{williams@mpia.de}

\author[0000-0002-0786-7307]{Thomas G. Williams}
\affiliation{\oxford}
\affiliation{\mpia}

\author[0000-0003-0378-4667]{Jiayi Sun}
\altaffiliation{CITA National Fellow}
\affiliation{\mcmaster}
\affiliation{\cita}

\author[0000-0003-0410-4504]{Ashley.~T.~Barnes}
\affiliation{\UBonn}

\author[0000-0002-3933-7677]{Eva Schinnerer}
\affiliation{\mpia}

\author[0000-0001-9656-7682]{Jonathan D. Henshaw}
\affiliation{Astrophysics Research Institute, Liverpool John Moores University, 146 Brownlow Hill, Liverpool L3 5RF, UK}
\affiliation{\mpia}

\author[0000-0002-6118-4048]{Sharon E. Meidt}
\affiliation{\ghent}

\author[0000-0002-0472-1011]{Miguel Querejeta}
\affiliation{\oan}

\author[0000-0002-7365-5791]{Elizabeth J. Watkins}
\affiliation{\ari}

\author[0000-0003-0166-9745]{Frank Bigiel}
\affiliation{\UBonn}

\author[0000-0003-4218-3944]{Guillermo A. Blanc}
\affiliation{The Observatories of the Carnegie Institution for Science, 813 Santa Barbara St., Pasadena, CA, USA}
\affiliation{Departamento de Astronom\'{i}a, Universidad de Chile, Camino del Observatorio 1515, Las Condes, Santiago, Chile}

\author[0000-0003-0946-6176]{M\'{e}d\'{e}ric~Boquien}
\affiliation{Centro de Astronomía (CITEVA), Universidad de Antofagasta, Avenida Angamos 601, Antofagasta, Chile}

\author[0000-0001-5301-1326]{Yixian Cao}
\affiliation{\mpe}

\author[0000-0002-5635-5180]{M\'elanie~Chevance}
\affiliation{\ita}
\affiliation{\cool}

\author[0000-0002-4755-118X]{Oleg V. Egorov}
\affiliation{\ari}

\author[0000-0002-6155-7166]{Eric Emsellem}
\affiliation{\eso}
\affiliation{\cral}

\author[0000-0001-6708-1317]{Simon C. O Glover}
\affiliation{\ita}

\author[0000-0002-3247-5321]{Kathryn Grasha}
\affiliation{\anu}

\author[0000-0002-8806-6308]{Hamid Hassani}
\affiliation{Department of Physics, University of Alberta, Edmonton, Alberta, T6G 2E1, Canada}

\author[0000-0002-4232-0200]{Sarah Jeffreson}
\affiliation{\cfa}

\author[0000-0002-9165-8080]{Mar\'ia J. Jim\'{e}nez-Donaire}
\affiliation{\oan}

\author[0000-0002-0432-6847]{Jaeyeon Kim}
\affiliation{\ita}

\author[0000-0002-0560-3172]{Ralf S.\ Klessen}
\affiliation{\ita} 
\affiliation{\iwr}

\author[0000-0001-6551-3091]{Kathryn Kreckel}
\affiliation{\ari}

\author[0000-0002-8804-0212]{J.~M.~Diederik~Kruijssen}
\affiliation{\cool}

\author[0000-0003-3917-6460]{Kirsten L. Larson}
\affiliation{AURA for the European Space Agency (ESA), Space Telescope Science Institute, 3700 San Martin Drive, Baltimore, MD 21218, USA}

\author[0000-0002-2545-1700]{Adam~K.~Leroy}
\affiliation{\OSU} 
\affiliation{\CCAPP}

\author[0000-0001-9773-7479]{Daizhong Liu}
\affiliation{\mpe}

\author[0000-0002-0873-5744]{Ismael Pessa}
\affiliation{\mpia}
\affiliation{Leibniz-Institut f\"{u}r Astrophysik Potsdam (AIP), An der Sternwarte 16, 14482 Potsdam, Germany}

\author[0000-0003-3061-6546]{J\'{e}r\^{o}me Pety}
\affiliation{IRAM, 300 rue de la Piscine, 38400 Saint Martin d'H\`eres, France}
\affiliation{LERMA, Observatoire de Paris, PSL Research University, CNRS, Sorbonne Universit\'es, 75014 Paris}

\author[0000-0001-5965-3530]{Francesca Pinna}
\affiliation{\mpia}

\author[0000-0002-5204-2259]{Erik Rosolowsky}
\affiliation{Department of Physics, University of Alberta, Edmonton, Alberta, T6G 2E1, Canada}

\author[0000-0002-4378-8534]{Karin M. Sandstrom}
\affiliation{Department of Physics, University of California, San Diego, 9500 Gilman Drive, San Diego, CA 92093, USA}

\author[0000-0002-0820-1814]{Rowan Smith}
\affiliation{\manch}

\author[0000-0001-6113-6241]{Mattia C. Sormani}
\affiliation{\ita}

\author[0000-0002-9333-387X]{Sophia Stuber}
\affiliation{\mpia}

\author[0000-0002-8528-7340]{David A. Thilker}
\affiliation{Department of Physics and Astronomy, The Johns Hopkins University, Baltimore, MD 21218, USA}

\author{Bradley C. Whitmore}
\affiliation{\stsci}



\begin{abstract}

We combine \textit{JWST} observations with ALMA CO and VLT-MUSE H$\alpha$ data to examine off-spiral arm star formation in the face-on, grand-design spiral galaxy NGC~628. We focus on the northern spiral arm, around a galactocentric radius of 3--4 kpc, and study two spurs. These form an interesting contrast, as one is CO-rich and one CO-poor, and they have a maximum azimuthal offset in MIRI 21$\mu$m and MUSE H$\alpha$ of around 40$^\circ$ (CO-rich) and 55$^\circ$ (CO-poor) from the spiral arm. The star formation rate is higher in the regions of the spurs near to spiral arms, but the star formation efficiency appears relatively constant. Given the spiral pattern speed and rotation curve of this galaxy and assuming material exiting the arms undergoes purely circular motion, these offsets would be reached in 100 -- 150 Myr, significantly longer than the 21$\mu$m and H$\alpha$ star formation timescales (both \textless10 Myr). The invariance of the star formation efficiency in the spurs versus the spiral arms indicates massive star formation is not only triggered in spiral arms, and cannot simply occur in the arms and then drift away from the wave pattern. These early {\it JWST} results show that in-situ star formation likely occurs in the spurs, and that the observed young stars are not simply the `leftovers' of stellar birth in the spiral arms. The excellent physical resolution and sensitivity that {\it JWST} can attain in nearby galaxies will well resolve individual star-forming regions and help us to better understand the earliest phases of star formation.

\end{abstract}

\keywords{galaxies: individual (NGC~628) --- galaxies: spiral --- galaxies: ISM --- galaxies: star formation}


\section{Introduction} \label{sec:intro}

\begin{figure*}[ht!]
\plotone{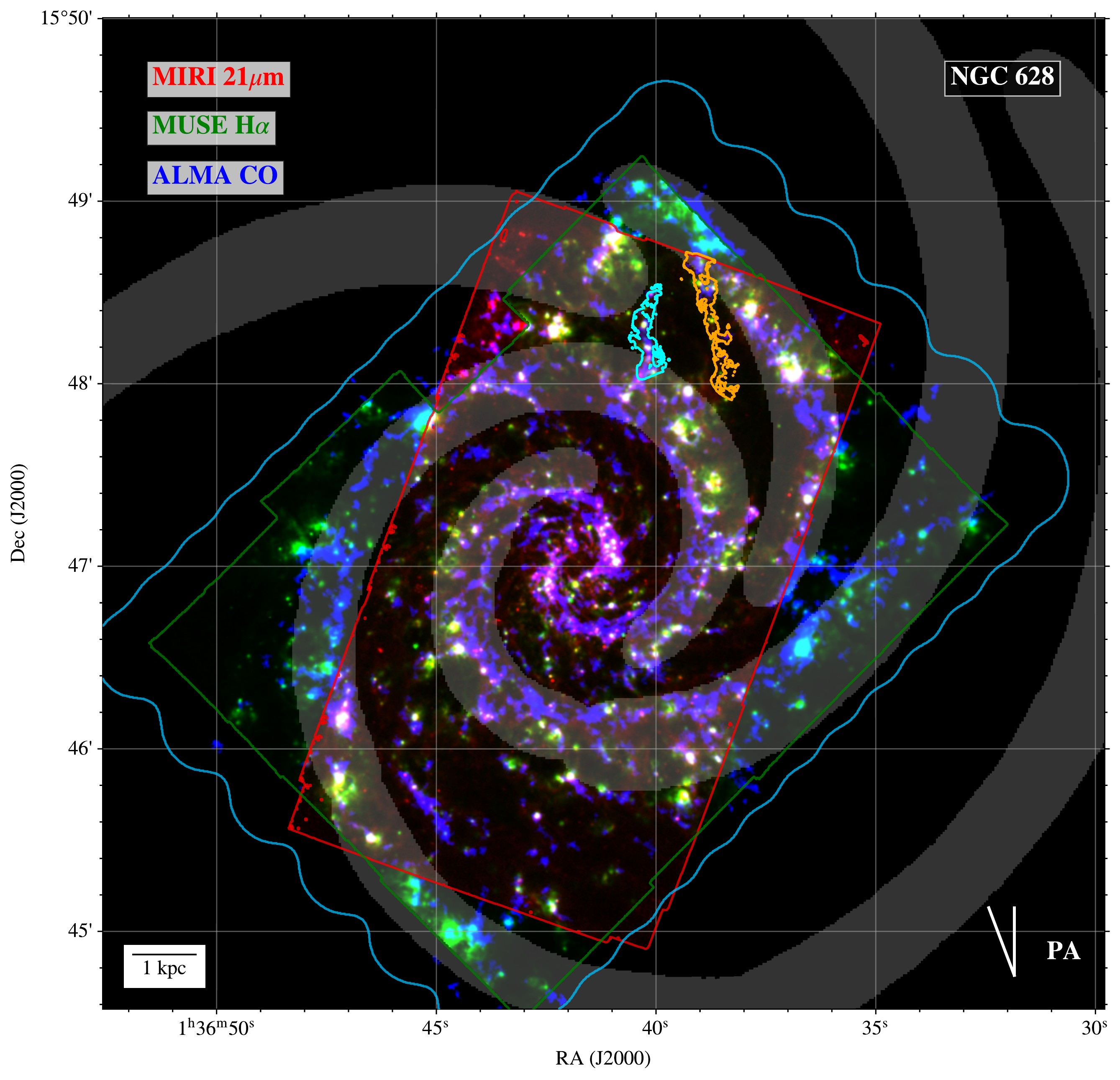}
\caption{Three-colour image of NGC~628, with ALMA CO in blue, MUSE H$\alpha$ in green, and {\it JWST} 21$\mu$m in red. The blue, red, and green boxes show the extent of each corresponding observation. The two spurs we focus on are shown as cyan (CO-rich), and orange (CO-poor) contours (see \S\ref{sec:spur_offset_timescale}). Also shown are the spiral arms from the environmental mask, in gray. The position angle (21$^\circ$; corresponding to $\theta = 0^\circ$ in Figure \ref{fig:polar_unwrap}) is indicated in the lower-right, and a 1~kpc scalebar is shown in the lower-left.}
\label{fig:jwst_muse_alma}
\end{figure*}


\begin{figure*}[ht!]
\plotone{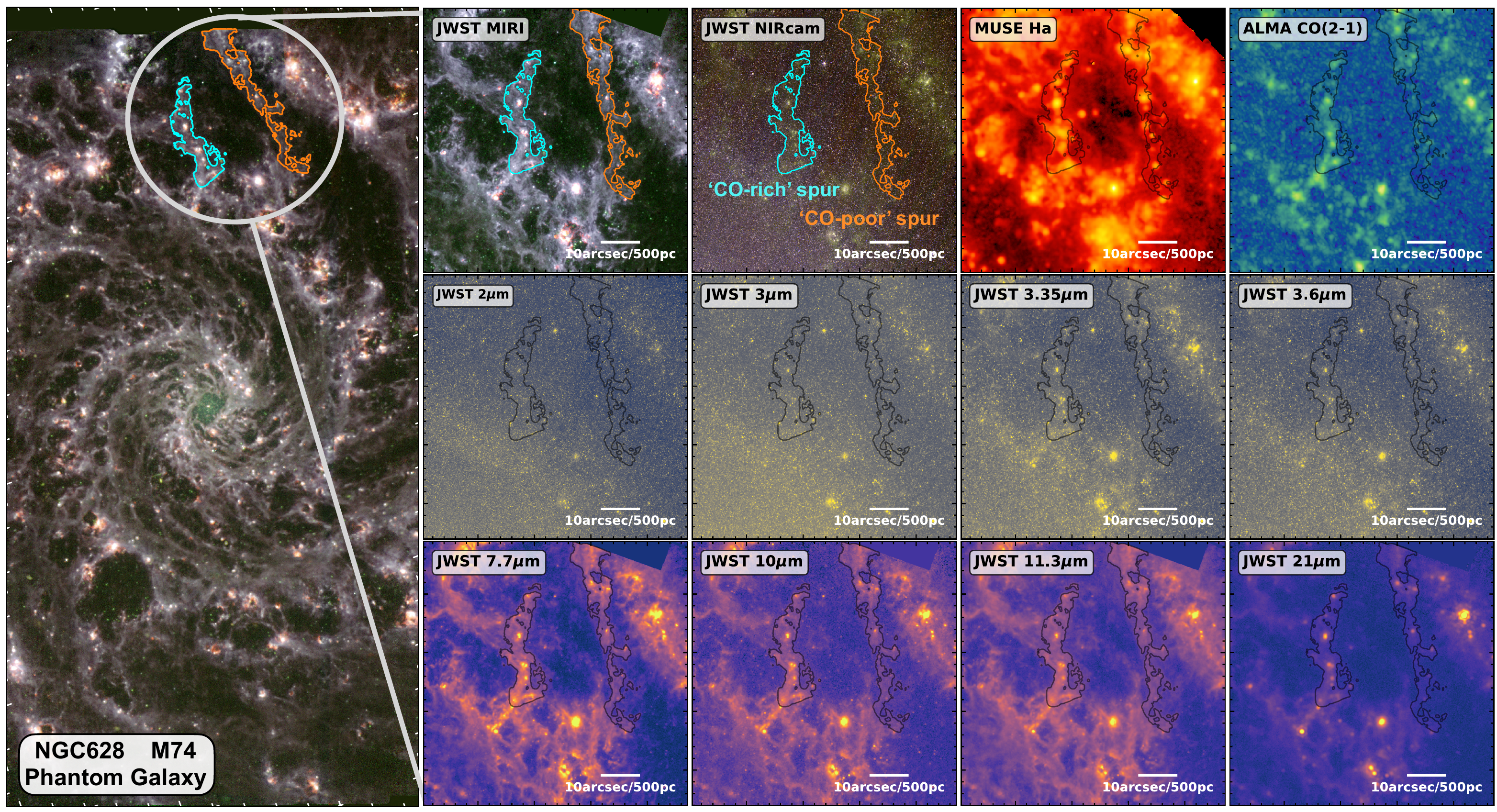}
\caption{A holistic overview of the spurs analysed in this study. {\it Left:} three-colour image of NGC628 produced from the 770W (blue), 1000W (green), and 1130W (red) band filters from the {\it JWST} \citep{2023Lee}, and overlaid in orange is the continuum subtracted {\it HST}-H$\alpha$, with CO-rich and CO-poor spur (see \S\ref{sec:spur_offset_timescale}) highlighted in cyan and orange, respectively. {\it Right, top row}: from left to right, three-colour MIRI (same as left panel) and NIRCam (red: 200W, green: 300M, blue: 335M) zooms of the spurs, as well as MUSE H$\alpha$ and ALMA CO. {\it Right, middle}: From left to right, increasing {\it JWST} NIRCam wavelengths, showing the stellar light. {\it Right, bottom}: From left to right, increasing {\it JWST} MIRI wavelengths, showing the ISM emission.}
\label{fig:spurmaps}
\end{figure*}

\begin{figure*}[ht!]
\plotone{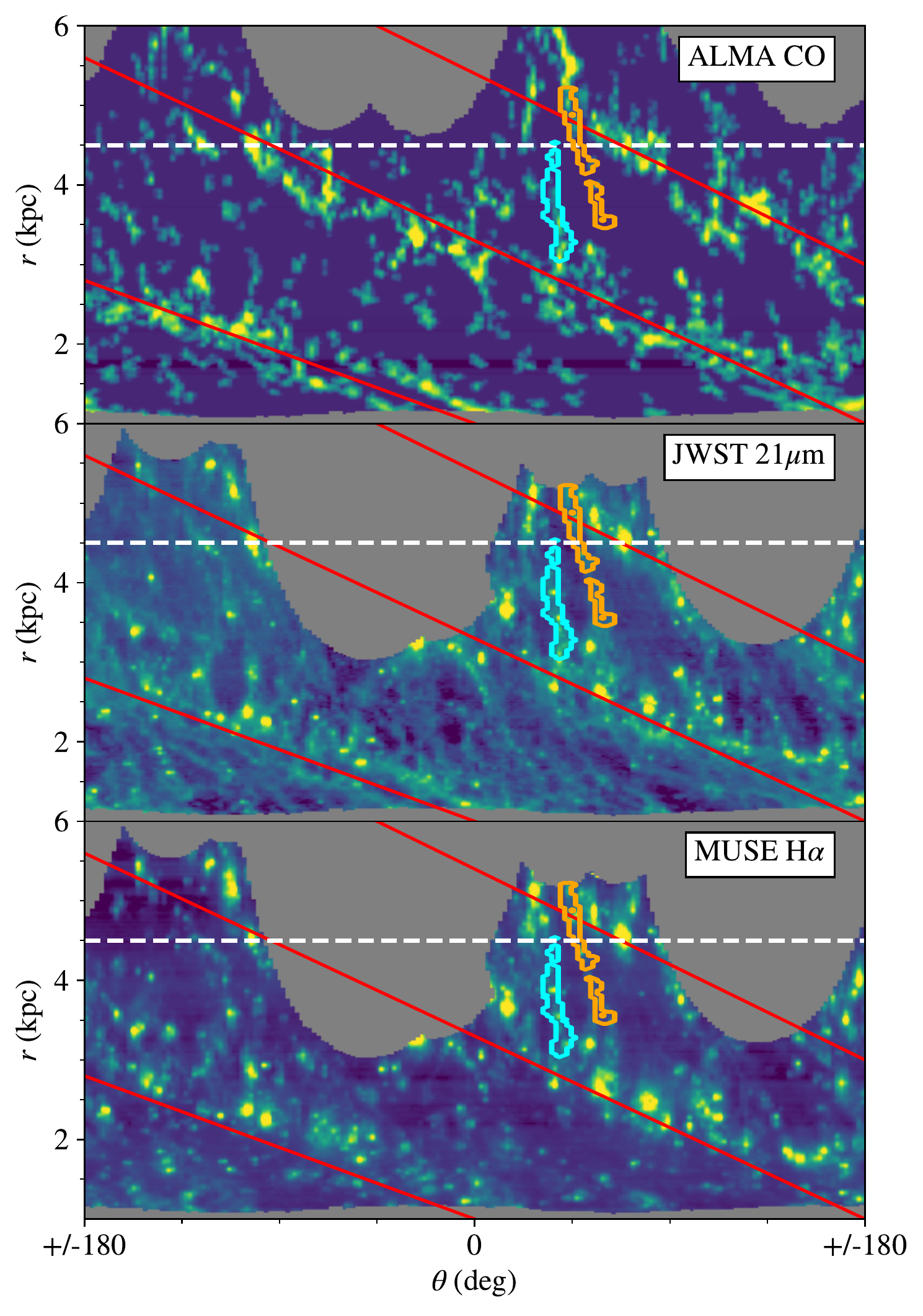}
\caption{Polar deprojection of NGC~628 in CO (top), 21$\mu$m (middle), and MUSE H$\alpha$ (bottom). $0^\circ$ is defined as the position angle (see Fig. \ref{fig:jwst_muse_alma}), and $\theta$ increases from left to right. The nominal co-rotation radius from \cite{2021Williams} is shown as a horizontal dashed white line. We show the approximate ridge of three spiral arms as red lines (determined from the CO). We note there is a clear offset between this CO ridge and the 21$\mu$m/H$\alpha$ ridges \citep[see also, e.g.][]{2018Kreckel} The cyan contour indicates the `CO-rich' spur, and the orange the `CO-poor' spur.}
\label{fig:polar_unwrap}
\end{figure*}

Spiral arms are a distinctive characteristic of star-forming galaxies, featuring large, curved arcs across the galaxy discs as gas is compressed and star formation occurs. Historically, spiral arms have been seen as the sites where the majority of stars form within galaxies \citep[e.g.][]{1953MorganWhitfordCode, 1969Roberts, 2013Louie}. This is thought to stem from high gas densities achieved in the spiral arms, combined with low shear \citep{2007Elmegreen}, which favours cloud (and from this, star) formation \citep[see reviews by][]{2014DobbsBaba, 2022Chevance}. Star formation triggering is also thought to take place in spiral arms, given the shocking that occurs at these locations \citep{1969Roberts}, and the potential for cloud-cloud collisions \citep[e.g.][]{1998Kennicutt, 2000Tan, 2014Longmore, 2021Fukui, 2020Chevance}. Here, gas compressed by molecular clouds colliding can lead to an episode of star formation.

Modern views of the process of star formation in spiral arms built from observations and simulations emphasise that the pattern of star formation in and around these locations is complex \citep[e.g.][]{2010DobbsPringle, 2017Chandar, 2017Schinnerer, 2020KimKimOstriker}. This is thought to reflect that spiral arms are not smooth, singular structures but instead themselves host a complex array of substructures. These structures have a variety of names, such as spurs or feathers \citep[see][for a discussion of the nomenclature]{2006LaVigne}. These features protrude from the spiral arms, are fairly regularly spaced with azimuth, and are predicted in simulations to extend to kpc scales. The origin of these spurs is currently unclear. Several mechanisms have been proposed, including gravitational instabilities \citep[e.g.][]{2006Dobbs}, magneto-Jeans instabilities \citep[e.g.][]{2006KimOstriker}, wiggle instabilities \citep[e.g.][]{2004WadaKoda,2022Mandowara}, supernova feedback, or formation on the edges of superbubbles \citep[primarily feedback driven expansions of gas; see e.g.][]{1997Oey, 2020KimKimOstriker}. Depending on the formation mechanism, these spurs are also thought to be viable sites of further fragmentation and collapse.  Thus, star formation may not occur exclusively in the high density spiral arm ridge. Measuring the star formation as a function of distance from spiral arms is critical for testing spur formation pathways, and can help us to better understand star formation associated with the spiral-arm passage of gas.

Spurs can be seen in molecular gas tracers \citep{2008Corder, 2009Koda, 2017Schinnerer, 2023Stuber}, as well as in the dust morphology \citep{2006LaVigne}. The goal of this work is to ask and answer whether stars are forming natively within spurs or if stars have formed in the dense spiral ridge and drifted to their present positions coincident with the spurs. Certainly for M~51, the former appears to be the case \citep{2017Schinnerer}, as typical extragalactic star formation rate tracers (H$\alpha$, 24$\mu$m) are coincident with CO in the spurs, rather than with the ridge of the CO spiral arm. Localizing the natal site of star formation requires the use of a tracer of the youngest, most embedded phase of the star formation process (i.e. with timescales \textless10~Myr). This ensures that we can catch star formation `in the act'. For this, the mid- and far-infrared are ideal, as bright and compact emission at these wavelengths directly traces the hot dust heated by young, embedded stars. However, given the limited resolution of the {\it Spitzer} MIPS \citep{2004Rieke} instrument (which had a resolution of $\sim$300~pc at a distance of 10~Mpc, but was the only viable instrument in this wavelength range before {\it JWST}), until now localising the mid infrared (MIR) emission to spurs or spiral arms has been challenging in galaxies outside the Local Group. This means establishing whether the phenomenon of star formation within spurs is unique to M~51 or a general feature of all disc galaxies is still an open question, with important implications for star formation models.

In this Letter, we use new {\it JWST} observations taken as part of the PHANGS-JWST Treasury Program \citep[PI J.~C.~Lee;][]{2023Lee} to study star formation in the spiral arms of NGC~628. We test whether star formation off of the spiral arms in NGC~628 could be from stars forming within spiral arms and then drifting, or whether stars are formed locally within spurs. The structure of this Letter is as follows: we briefly describe why NGC~628 is an ideal target for this study, and the data provenance in \S\ref{sec:data}, identify our spiral arm region of interest and the timescales for offset between spiral arm and spur in \S\ref{sec:spur_offset_timescale}. We conclude in \S\ref{sec:conclusions}.

\section{NGC~628 and Data} \label{sec:data}

As an archetypal grand-design spiral galaxy, NGC~628 is an ideal target for studies of spiral arms, given its clear arm structure and lack of a bar. Located at a distance of 9.84~Mpc \citep{2017McQuinn, 2021Anand, 2021bAnand}, NGC~628 is almost face-on \citep[$i=8.9^\circ$;][]{2020Lang}, and aligned nearly north-up with a position angle of $20.7^\circ$ \citep{2020Lang}. It is also the only galaxy in PHANGS--MUSE with a robustly measured spiral arm pattern speed from stellar kinematics by \citet[][the rest of the pattern speeds in this work being attributed to bars]{2021Williams}, which is necessary to obtain timescales for the spur offset (\S\ref{sec:spur_offset_timescale}). We present a three-colour composite image in Figure \ref{fig:jwst_muse_alma} including the data we will use in this study, and Figure \ref{fig:spurmaps} presents a more holistic overview of the spurs, which shows the wealth of high-quality observations, and rich detail present in the PHANGS (and especially the PHANGS-JWST) data that exists for this galaxy. Returning to Figure \ref{fig:jwst_muse_alma}, in blue, we show a CO(${\it J}= 2-1$), hereafter CO, moment 0 map from ALMA \citep{2021aLeroy, 2021bLeroy}, tracing the cold molecular gas across the galactic disc. Here, a number of gas spurs are visible as structures that are almost perpendicular to the spiral arm. These data have a resolution of $\sim$1\arcsec ($\sim$50~pc) and a sensitivity of $\sim$1~K~km~s$^{-1}$. In green, we show H$\alpha$ emission from VLT-MUSE observations as part of PHANGS-MUSE \citep{2022Emsellem}, tracing young stars that are producing ionising radiation but have blown a hole in their natal cloud. The MUSE data also have a resolution of around 1\arcsec, with an H$\alpha$ sensitivity of $\sim1.5 \times 10^{37}$~erg~s$^{-1}$~kpc$^{-2}$. H$\alpha$ emission traces star formation over timescales of \textless10~Myr \citep[e.g.][]{2006Moustakas, 2012Leroy, 2012KennicuttEvans, 2014Boquien}. In red, we show 21$\mu$m {\it JWST} data \citep{2023Lee}, with a resolution of 0.67\arcsec\ and a surface brightness sensitivity of around 0.3~MJy~sr$^{-1}$. In star-forming regions, this wavelength traces young, highly embedded star formation \citep[see, e.g., radiative transfer models by][]{2014DeLooze, 2019Williams}, with an emitting timescale in the star-forming regions of $10$\,Myr for NGC~628 \citep{2023Kim}. Finally, we overlay spiral arms as defined by \cite{2021Querejeta} in gray, based on {\it Spitzer} 3.6$\mu$m imaging, which traces the spiral arms from the old stars.

\section{Spur Offset and Timescale}\label{sec:spur_offset_timescale}

\begin{figure*}[ht!]
\plotone{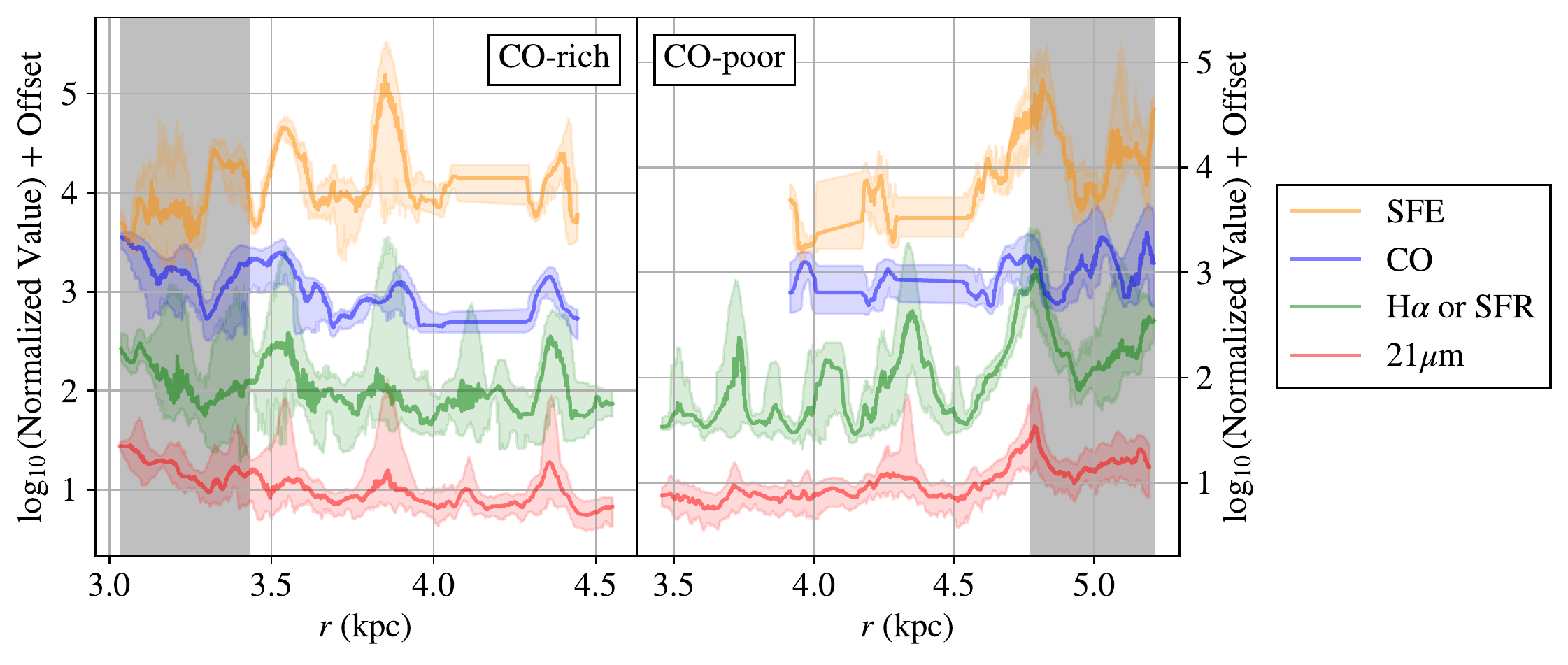}
\caption{Profiles of 21$\mu$m, Balmer-corrected H$\alpha$ (equivalent to SFR), CO, and SFE values across the CO-rich and CO-poor spurs (the mask generated from the contours in Fig. \ref{fig:jwst_muse_alma}) as a function of galactocentric radius. The solid line shows the rolling median of the data, and the shaded regions the 16$^{\rm th}$ and 84$^{\rm th}$ percentiles. Each intensity is normalised by its 50$^{\rm th}$ percentile value within the spur mask, and is offset for visual clarity. The shaded gray region indicates the parts of the spur within the spiral arm mask. For reference, the offsets are 1, 2, 3, and 4 for 21$\mu$m, SFR, CO, and SFE, respectively. These lines correspond to the median value of 1.6MJy~sr$^{-1}$ (21$\mu$m), $5.3\times10^{-3}$M$_\odot$~kpc$^{-2}$ \citep[SFR surface density, assuming a][conversion factor]{2007Calzetti}, 4.7K~km~s$^{-1}$ (CO), and $3.4\times10^{-7}$yr$^{-1}$ (SFE) for the CO-rich spur, and 1.2MJy~sr$^{-1}$, $4.2\times10^{-3}$M$_\odot$~kpc$^{-2}$, 3.7K~km~s$^{-1}$, and $3.3\times10^{-7}$yr$^{-1}$ for the CO-poor spur.}
\label{fig:radial_sfr_sfe}
\end{figure*}

\begin{figure}[ht!]
\plotone{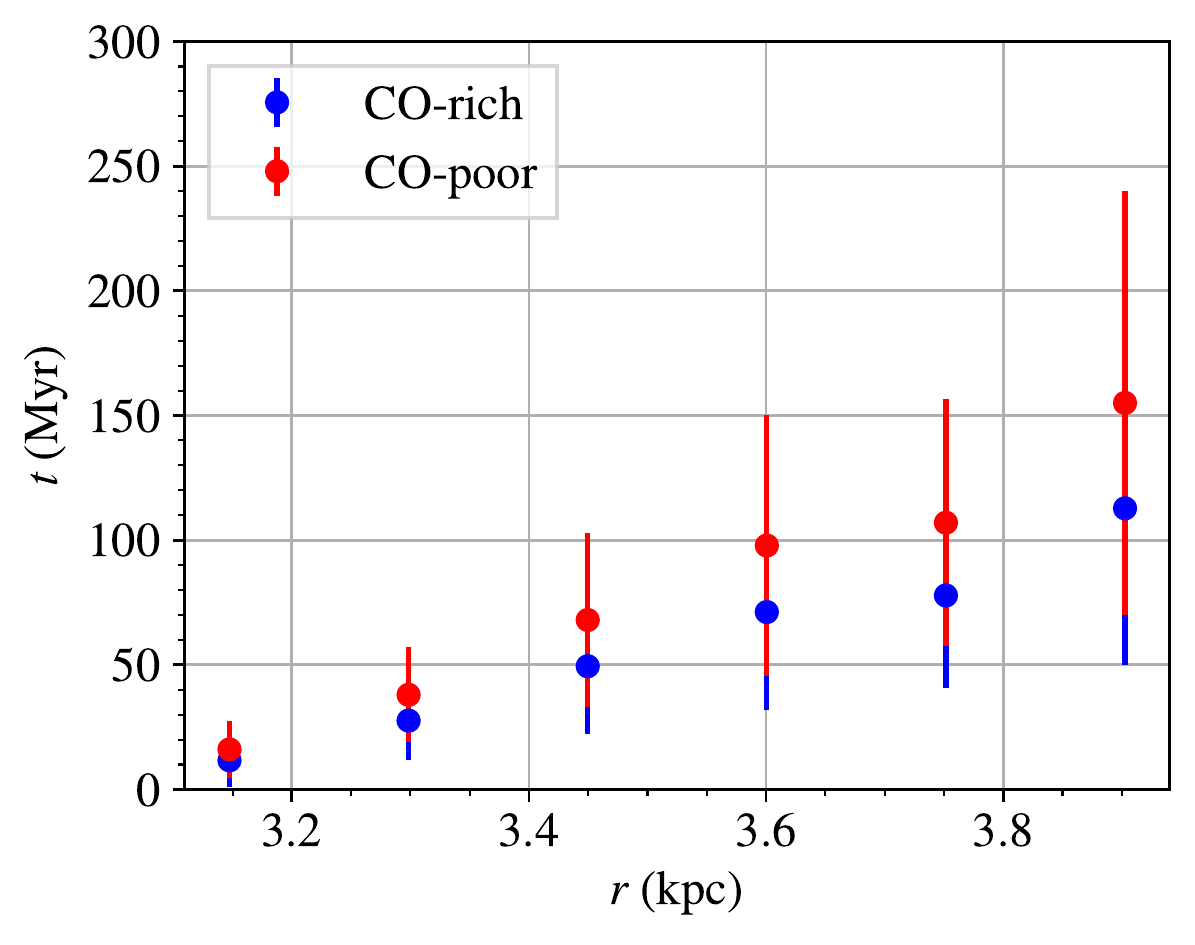}
\caption{Offset timescale (see Eq. \ref{eq:timescale}) as a function of galactocentric radius for the CO-rich (blue) and CO-poor (red) spurs, calculated between a galactocentric radius of 3 and 4~kpc.}
\label{fig:timescales}
\end{figure}

In Figure \ref{fig:polar_unwrap}, we perform a polar (i.e.\ $r$, $\theta$ space) remapping (i.e.\ deprojection and derotation) of Figure \ref{fig:jwst_muse_alma}. Here, 0$^\circ$ corresponds to the position angle of the galaxy shown in Figure \ref{fig:jwst_muse_alma}, with $\theta$ increasing in a clockwise direction. In this polar projection, the spiral arms appear as nearly straight lines (these are well-described as log-spirals in \citealt{2021Querejeta}, but the difference here is minor), and the spurs off the arms become more clear. There is also a clear offset between these three tracers in the spiral arms, as a consequence of the spiral pattern \citep[e.g.][]{2009Egusa, 2018Kreckel}. This is not a focus of this paper, but we note that the high resolution now available with these tracers may be useful for future direct measurements of spiral pattern speeds. There is a significant amount of CO, H$\alpha$, and 21$\mu$m emission in spurs off the spiral arms, which is visible off all the spiral arms in Figure \ref{fig:jwst_muse_alma}. 

We will focus on two spurs in particular, both between a (deprojected) galactocentric radius of 3 to 4~kpc. These spurs are present near co-rotation \citep{2021Williams}, implying the drift times from the spiral arm will be quite long. The first has a maximum offset from the spiral arm of $\theta \simeq 40^\circ$, and has clearly associated CO, H$\alpha$ and 21$\mu$m emission. We will refer to this as the `CO-rich' spur. The second has a maximum offset from the spiral arm of $\theta \simeq 55^\circ$, and is well detected in H$\alpha$ and 21$\mu$m but has no associated emission in the `strict mask' (i.e. a high-confidence, but lower completeness mask, see \citealt{2021aLeroy} for details) CO moment 0 image (shown in Figure \ref{fig:jwst_muse_alma}, although it is barely detected in the `broad mask' moment 0, which has lower confidence but higher completeness as compared to the strict mask, and is not shown here, but see \citealt{2021aLeroy}). Both of these spurs are also clearly detected in the other MIRI (7.7, 10, and 11.3$\mu$m) bands, and the emission is coincident with that at 21$\mu$m. We have selected these two as a test case, as they are neighbouring spurs but quite different in their ISM composition. We reserve a more thorough cataloguing and study of spurs for future work with the larger PHANGS-JWST sample. 

The total 21$\mu$m flux outside the spiral arms as defined by the environmental mask (excluding the central 1.2~kpc diameter region based on photometric decomposition by \citealt{2015Salo}, where disentangling the spiral arms from any potential stellar bulge or nuclear component is difficult) is around 60\%, indicating a non-negligible amount of star formation outside the spiral arms. We caution that the environmental mask is defined by {\it Spitzer} data and the gas and stellar spiral arms may not necessarily coincide. However, the spiral arm width follows an empirical definition based on the CO emission, to attempt to overcome this \citep{2021Querejeta}. There also may be a significant amount of diffuse emission at 21$\mu$m flux, that is unlikely to originate from star formation \citep{2023Leroy}. In this sense, the percentage is likely an upper limit to the amount of 21$\mu$m flux that can be ascribed to star-formation. 

We next investigate how the flux profiles of the CO, H$\alpha$ and 21$\mu$m vary with galactocentric radius along these two spurs, to better understand the role the spiral arms have in enhancing the star formation rate (SFR) and star formation efficiency (SFE; SFR per unit molecular gas mass). Using the spur contour in Figure \ref{fig:jwst_muse_alma} as a mask, we calculate radial profiles of the intensity of the three tracers for the two spurs. We use Balmer decrement-corrected H$\alpha$ as a proxy for the SFR, and also calculate the profile for SFE (i.e. corrected H$\alpha$/CO). We show the profiles in Figure \ref{fig:radial_sfr_sfe}. Between the spiral arms, the SFR appears to be relatively constant, arguing against the idea of an evolutionary sequence with stars further out in the spurs being formed at an early time to those closer to the spiral arm. However, there is an increase towards the spiral arms -- the inner arm for the CO-rich spur, and the outer for the CO-poor. This agrees with simulations showing that the spiral arms act to concentrate star forming regions, leading to an overall increase in the SFR surface density \citep[e.g.][]{2020KimKimOstriker}. Yet, comparing the 21$\mu$m and H$\alpha$ to that of the CO, we see a good correspondence between the profiles, with the tracers tending to upturn at the same radii. Indeed, the SFE profile bears this out -- the SFE sometimes shows strong variation along the spurs, but the increases in SFE are localised and do not correlate with the spiral arm positions (although there is a slight increase towards the outer spiral arm in the case of the CO-poor spur). Taken together, these results advocate that the spiral arms gather together gas and star forming regions, but have little impact on how efficiently stars are being formed, as seen in larger (but lower resolution) samples \citep{2021Querejeta} or in simulations \citep[e.g.][]{2015Dobbs}.

We estimate the timescale for both of these features to appear, assuming they have drifted from the spiral arm with the passage of the density wave and at the same pattern speed. Following \cite{2009Egusa} we compute the timescale required for this spiral arm offset to occur (neglecting any non-circular motion), as
\begin{equation}\label{eq:timescale}
    t = 76.8\, {\rm Myr}\, \left(\frac{\Delta \theta}{45^\circ} \right) \left(\frac{\Omega(r) - \Omega_P}{10 \, {\rm km\, s^{-1} \, kpc^{-1}}} \right)^{-1} ,
\end{equation}
where $\Omega$ is the angular rotation velocity, $\Omega_P$ the pattern speed (both in km~s$^{-1}$~kpc$^{-1}$), $\Delta \theta$ the offset in degrees (i.e. the distance from spiral arm to spur along the $x$-axis in Fig. \ref{fig:polar_unwrap}, and will vary from 0 where the spur meets the spiral arm to some maximum offset), and $t$ the timescale in units of Myr. We use $\Omega_P = 31.1^{+4.0}_{-2.9}~{\rm km~s^{-1}~kpc^{-1}}$ from \cite{2021Williams}. This is a conservative value, as, for example, gravity will act to pull the gas back towards the spiral arm, lengthening the timescales. We obtain $\Omega(r)$ from the measured rotation curves in \citet{2020Lang}, which vary from around $36\,{
\rm km~s^{-1}~kpc^{-1}}$ to $39\,{
\rm km~s^{-1}~kpc^{-1}}$. We estimate the maximum spur offset from Figure \ref{fig:polar_unwrap}, and assume it varies linearly (as the spurs are mostly vertical in the polar projection) with $r$ up to a maximum offset at a galactocentric radius of 4~kpc. We assign a relatively conservative uncertainty to these values of 5$^\circ$.

The calculated timescales are shown in Figure \ref{fig:timescales}, and are indeed quite long, as expected. These numbers are also likely lower bounds, as processes like gravitational attraction towards to spiral arm ridge will only serve to make these timescales longer. We see that the values range from close to 0 at the point where the spur joins the spiral arm up to more than 100~Myr at the farthest extent of the spur, significantly longer than the timescale we would expect the H$\alpha$ and 21$\mu$m emission to be visible for, if star formation was initiated in the arms (\textless10~Myr, see \S\ref{sec:data}). The same conclusion was found by \cite{2017Schinnerer} in M~51, perhaps indicating that this is a general result within galaxies. Altogether, our analysis suggests that stars can form in-situ within spurs, rather than moving from the spiral arms. This has been seen in some recent simulations \citep[e.g.][]{2020Smith, 2021Tress}, and combined with results showing the star formation efficiency may not be higher in the spiral arms \citep[e.g.][]{2018Ragan,2021Querejeta}, these results point towards a picture where the spiral arms merely gather gas together, rather than being instrumental in causing the onset of star formation.

The fact that one of these spurs is rich in CO and the other poor is also intriguing, given their close proximity. It seems possible that these spurs could potentially be forming from superbubble expansion \citep{2020KimKimOstriker}, as these spurs are on the edge of one of the large bubbles catalogued in \cite{2023Barnes} and \citet{2023Watkins}, and so should be in roughly the same evolutionary state. Perhaps, then, some feedback mechanisms have been more efficient at destroying gas in one spur compared to the other, or maybe the CO-poor one is older. This could be addressed both by observing the coincidence of spurs and bubbles, and using stellar clusters from combined {\it HST/JWST} observations as `clocks'. This is beyond the scope of this work, but would be an interesting future study with a full PHANGS-JWST sample of spurs, bubbles, and stellar clusters.

\section{Conclusions}\label{sec:conclusions}

In this Letter, we have combined ALMA, VLT-MUSE and new {\it JWST} observations in the context of the PHANGS collaboration to examine the youngest, highly embedded stage of star formation in a CO-rich and CO-poor spur off the prominent northern spiral arm of NGC~628. These were chosen as a test case, as they are next to each other but clearly quite different in their ISM composition. Both of these spurs show an increase in star formation towards spiral arms, but little indication of an increase in the star formation efficiency. Given the angular offset of these spurs, assuming they are formed on the arm and drifted off due to the difference between circular rotation speed and arm pattern speed, we infer a timescale of around 100~Myr or more, an order of magnitude higher than the timescales of the H$\alpha$ and 21$\mu$m emission \citep{2021Kim, 2023Kim}. These results imply that stars are forming in-situ within the spurs, rather than being produced within the spiral arms and then travelling there.

This work represents an initial exploration into how {\it JWST} observations will redefine our view of the earliest phases of galactic-scale star formation, and how this affects the structure of the ISM and the process of star formation in different environments. In particular, combining a spur catalogue with both exposed (measured from {
\it HST}) and embedded (measured from {\it JWST}) stellar clusters will help to understand the evolutionary sequence of the structure of the ISM \citep[e.g.][]{2017Chandar}. With the full 19 galaxies of the PHANGS-JWST sample, we will be able to form a new picture of the highly complex, filamentary nature of the ISM.

\section*{Acknowledgments}
The authors would like to thank the anonymous referee for their constructive comments, which have improved this manuscript. TGW would also like to thank David Williams, for everything over the years. This work was carried out as part of the PHANGS collaboration. The analysis scripts underlying this work are available at \url{https://github.com/thomaswilliamsastro/jwst_ngc628}. All the {\it JWST} data used in this paper can be found in MAST: \dataset[10.17909/436y-rd76]{http://dx.doi.org/10.17909/436y-rd76}.

This work is based on observations made with the NASA/ESA/CSA JWST. The data were obtained from the Mikulski Archive for Space Telescopes at the Space Telescope Science Institute, which is operated by the Association of Universities for Research in Astronomy, Inc., under NASA contract NAS 5-03127. The observations are associated with JWST program 2107.
Based on observations collected at the European Southern Observatory under ESO programmes 094.C-0623 (PI: Kreckel), 095.C-0473,  098.C-0484 (PI: Blanc), 1100.B-0651 (PHANGS-MUSE; PI: Schinnerer), as well as 094.B-0321 (MAGNUM; PI: Marconi), 099.B-0242, 0100.B-0116, 098.B-0551 (MAD; PI: Carollo) and 097.B-0640 (TIMER; PI: Gadotti). This paper makes use of the following ALMA data:
ADS/JAO.ALMA\#2012.1.00650.S.
ALMA is a partnership of ESO (representing its member states), NSF (USA) and NINS (Japan), together with NRC (Canada), MOST and ASIAA (Taiwan), and KASI (Republic of Korea), in cooperation with the Republic of Chile. The Joint ALMA Observatory is operated by ESO, AUI/NRAO and NAOJ.
 
TGW and ES acknowledge funding from the European Research Council (ERC) under the European Union’s Horizon 2020 research and innovation programme (grant agreement No. 694343).
JS acknowledges the support of the Natural Sciences and Engineering Research Council of Canada (NSERC) through a Canadian Institute for Theoretical Astrophysics (CITA) National Fellowship.
JMDK gratefully acknowledges funding from ERC  via the ERC Starting Grant ``MUSTANG'' (grant agreement number 714907). COOL Research DAO is a Decentralized Autonomous Organization supporting research in astrophysics aimed at uncovering our cosmic origins.
JPe acknowledges support by the DAOISM grant ANR-21-CE31-0010 and by the Programme National ``Physique et Chimie du Milieu Interstellaire'' (PCMI) of CNRS/INSU with INC/INP, co-funded by CEA and CNES.
MC gratefully acknowledges funding from the Deutsche Forschungsgemeinschaft (DFG) through an Emmy Noether Research Group (grant number CH2137/1-1).
MB acknowledges support from FONDECYT regular grant 1211000 and by the ANID BASAL project FB210003.
EJW. RSK, and SCOG acknowledge funding from DFG via the Collaborative Research Center ``The Milky Way System''(SFB 881, funding ID 138713538, subprojects A1, B1, B2, B8, and P1).
KK, OE gratefully acknowledge funding from DFG in the form of an Emmy Noether Research Group (grant number KR4598/2-1, PI Kreckel).
FB would like to acknowledge funding from ERC via the ERC Consolidator Grant ``Empire'' (grant agreement No.726384).
JK gratefully acknowledges funding from DFG through the DFG Sachbeihilfe (grant number KR4801/2-1).
ER acknowledges the support of the Natural Sciences and Engineering Research Council of Canada (NSERC), funding reference number RGPIN-2022-03499.
RSK and SCOG acknowledge support from ERC via the ERC Synergy Grant ``ECOGAL'' (project ID 855130) and from the Heidelberg Cluster of Excellence (EXC 2181 - 390900948) ``STRUCTURES'', funded by the German Excellence Strategy. RSK also thanks the German Ministry for Economic Affairs and Climate Action for funding in  project ``MAINN'' (funding ID 50OO2206).
MQ acknowledges support from the Spanish grant PID2019-106027GA-C44, funded by MCIN/AEI/10.13039/501100011033.
KG is supported by the Australian Research Council through the Discovery Early Career Researcher Award (DECRA) Fellowship DE220100766 funded by the Australian Government. 
AKL gratefully acknowledges support by grants 1653300 and 2205628 from the National Science Foundation, by award JWST-GO-02107.009-A, and by a Humboldt Research Award from the Alexander von Humboldt Foundation.
G.A.B. acknowledges the support from ANID Basal project FB210003.
SKS acknowledges financial support from the German Research Foundation (DFG) via Sino-German research grant SCHI 536/11-1.

%

\vspace{5mm}
\facilities{JWST, ALMA, VLT-MUSE}


\software{astropy \citep{astropy:2013, astropy:2018},  
          numpy \citep{harris2020array},
          scipy \citep{2020SciPy-NMeth},
          scikit-image \citep{van2014scikit},
          matplotlib \citep{Hunter:2007},
          uncertainties\footnote{Uncertainties: a Python package for calculations with uncertainties, Eric O. LEBIGOT, \url{http://pythonhosted.org/uncertainties/}}
          }






\bibliography{bibliography}{}
\bibliographystyle{aasjournal}

\suppressAffiliationsfalse
\allauthors


\end{document}